\documentstyle[aps,graphicx,tighten,preprint]{revtex}

\begin{document}
\title{New Tests of 
Macroscopic Local Realism using Continuous Variable Measurements}
 \vskip 1 truecm
\author{M. D. Reid\\
}
\address{Physics Department, University of Queensland, Brisbane, Australia\\ }  	  
\date{\today}
\maketitle
\vskip 1 truecm
\begin{abstract}
We show that quantum mechanics predicts an Einstein-Podolsky-Rosen 
paradox (EPR), and also a contradiction with local 
hidden variable 
theories, for photon number measurements which have limited resolving 
power, to the point of 
imposing an uncertainty in the photon number result which is 
macroscopic in absolute terms. We  
show how this can be interpreted as a failure of a new, very strong premise, 
called macroscopic 
local realism. We link this premise to the Schrodinger-cat paradox. Our proposed experiments ensure all fields incident on each  
measurement apparatus are macroscopic. We show that an alternative measurement 
scheme corresponds to balanced homodyne detection of quadrature phase 
amplitudes. The implication is that where either EPR correlations or   
failure of local realism is 
predicted for quadrature phase amplitude measurements, one can 
potentially perform a modified experiment which would lead to   
conclusions about the much stronger premise of macroscopic local 
realism. 
\end{abstract}
\narrowtext
\vskip 0.5 truecm
\section{Introduction}

In 1935 Einstein, Podolsky and Rosen~\cite{1} (EPR) formulated
an argument, now experimentally realised~\cite{2}, in an attempt to show that quantum mechanics is an incomplete
theory. The EPR argument is based on the premise of 
local realism. Bell~\cite{3} in 1966 showed that the premise of local 
realism (local hidden variable theories) was incompatible quantum mechanics.  
``Local realism'' has now been 
essentially disproved by experiments~\cite{4} 
based on Bell's theorem or those of Greenberger, 
Horne and Zeilinger (GHZ)~\cite{3}.

To date the EPR 
and Bell theorems and experimental efforts 
primarily focus on measurements intrinsically 
microscopic, in that one requires to clearly distinguish 
between results (eigenvalues of the appropriate quantum operator) 
which are microscopically distinct. (An exception to this is the work 
of Peres~\cite{18}.)
Previous results~\cite{5} have indicated failure of 
 local realism for macroscopic systems, but the violations are still 
 apparently only indicated where measurements must resolve 
 microscopically different 
 results, such as adjacent photon number or spin values. It
  is not clear whether one 
 is testing a premise different to that tested in the microscopic 
 experiments.

We propose a strategy for testing local realism at a 
macroscopic level, in the sense emphasised by Schrodinger~\cite{6,7} in his 
famous ``cat'' paradox and also by Leggett and Garg~\cite{8}, where one 
considers macroscopically distinct outcomes.  We define in section 2 the 
premise of 
``macroscopic local realism''~\cite{9}, in such a way that it relates to the 
Schrodinger-cat example of macroscopically distinct states.

 In section 3 we present an EPR 
 argument based on
the validity of ``macroscopic local realism'',
which has not yet been questioned. We suggest 
that modifications to an experiment already performed by Ou et al~\cite{2} 
would realise this macroscopic EPR argument and would leave no logical 
alternative except to deny the validity of macroscopic local realism 
or else to accept the incompleteness of quantum mechanics, in the 
sense proposed by EPR.

In section 4 we present a quantum state which allows a violation of a 
Bell-inequality even where all uncertainties in measurements are 
macroscopic, and show how this implies a predicted failure of 
macroscopic local realism. 

  \section{Definition of Macroscopic Local Realism}

 In 1935 
 Einstein, Podolsky and Rosen~\cite{1} defined ``local realism''.
  ``Realism'' implies 
 that if one can predict with 
 certainty the result of a measurement of a physical quantity 
 at $A$, without disturbing the 
 system $A$, then the results of the measurement were predetermined.  
 One has an 
 ``element of reality'' corresponding to this 
 physical quantity. The element of reality is a variable which assumes 
 one of a set of values which are the predicted results of the 
 measurement. This value gives the result of the 
 measurement, should it be performed. 
 Locality states that events at $A$ cannot, 
 instantaneously, disturb a second system at $B$ 
 spatially separated from $A$. Taken together   
 ``local realism''  
 implies that, if one can predict the result of a measurement of a 
 physical quantity at $A$, by 
 making a simultaneous measurement at $B$, then the result of the 
 measurement at $A$ is described by an element of reality.
 
EPR assumed quantum mechanics to be correct in
predicting the existence of two spatially separated particles with correlated positions, and also correlated
momenta. The key quantity in
establishing the EPR argument is the precision (call the associated 
error $\Delta )$ to which
the result of the potential position measurement at $A$ can be inferred by the
measurement at $B$. This specifies an associated indeterminacy (error 
$\Delta )$ in
the ``element of reality'' $x$. In the original EPR gedanken example, 
$\Delta $ is zero. ``Local realism'' establishes two 
``elements of reality'', $x$ and $p$ 
which simultaneously exist to give precisely the
result of a potential position or momentum measurement, respectively. No
description of this nature exists within quantum mechanics, since any
quantum wavefunction gives an indeterminacy $\Delta x$ and $\Delta p$ in
position and momentum respectively, in accordance with the uncertainty
relation $\Delta x  \Delta p \geq \hbar /2$. In this way, the EPR
argument, based on the validity of ``local realism'', allows one to conclude
that quantum mechanics is incomplete.

Macroscopic local realism~\cite{9} is defined 
   as a premise stating the following. If one can predict 
   the result of a measurement at $A$ by performing 
 a simultaneous measurement on a spatially separated system $B$, then the 
 result of the measurement at $A$ is predetermined but 
 described by an element of reality 
 which has an indeterminacy in each of its possible  values, so 
 that only values macroscopically 
 different to those predicted are excluded.
 
 Macroscopic local realism 
 is based on a ``macroscopic locality'', 
 which states that measurements at a 
location $B$ cannot instantaneously induce macroscopic 
changes (for example the dead to alive state of a cat, or a change 
between macroscopically 
different photon numbers) in a second system $A$ spatially 
separated from $B$. 
 Macroscopic local realism also incorporates a 
 ``macroscopic realism'', since it implies elements of reality with (up 
 to) a macroscopic indeterminacy. 
 Suppose 
our ``Schrodinger's cat~\cite{6}'' is correlated with a second 
spatially separated system, for 
example a gun used to kill the cat.  The strength of 
macroscopic local realism is understood when one realises that its 
rejection in this example means we cannot think of the cat as being 
either dead or alive, even though we can predict the dead or alive 
result of ``measuring'' the cat, without disturbing the cat, 
by measurement on the correlated spatially-separated second system. 

 \section{An EPR Argument based on Macroscopic Local Realism}

 We consider a new EPR situation, depicted in Figure 1, where 
 uncertainties in ``elements of reality''
become macroscopic. The  $\hat a_{\pm}$ and $\hat b_{\pm}$ are boson operators for four
  fields, described by the quantum state $|\psi\rangle$. 
 Fields  $\hat a_{\pm}$ and $\hat b_{\pm}$ propagate 
 towards the spatially separated locations $A$ 
 and $B$ respectively. We measure simultaneously at $A$ and $B$ the Schwinger spin operators 
  \begin{eqnarray}
     \hat S_{\theta}^A &=& \hat S_{x}^{A}¥\cos{\theta} +
     \hat S_{y}^{A}¥\sin{\theta} \nonumber \\
    &=&(\hat a_+^\dagger  \hat a_- \exp (-i\theta)+\hat a_+  \hat 
a_-^\dagger \exp 
(i\theta))/2
\end{eqnarray} 
       and 
       \begin{eqnarray}
      \hat S_{\phi}^{B} &=& \hat S_{x}^{B}¥\cos{\phi} +
      \hat S_{y}^{B}¥\sin{\phi} \nonumber \\
          &=&(\hat b_+^\dagger  \hat b_- \exp (-i\phi)+\hat b_+  \hat 
b_-^\dagger \exp 
(i\phi))/2
      \end{eqnarray} 
       respectively, where  
   $\hat S_{x}^{A}=(\hat a_+^\dagger \hat a_- + \hat a_-^\dagger \hat a_+)/2, 
  \hat S_{y}^{A}¥=(\hat a_+^\dagger \hat a_- - \hat a_-^\dagger \hat a_+)/2i$ and 
  $\hat S_{z}^{A}¥=(\hat a_+^\dagger 
  \hat a_+-\hat a_-^\dagger \hat a_-)/2$, and similarly for the modes at $B$.

We propose to measure, at $A$,  
 $\hat S_x^A$ or $\hat S_y^A$, by selecting $\theta=0$ or 
 $\theta=\pi/2$. At $B$ the measurement is either  
$S_x^B$ or $S_y^B$. 
In Figure 1 the measurement at $A$ is performed by first  
mixing $\hat a_{\pm}$ using phase shifts and  
beam splitters to give two new fields 
 $\hat a_{-}^{'}=(\hat a_{-}-\hat a_{+})/\sqrt{2}$ and 
    $\hat a_{+}^{'}=i(\hat a_{-}+\hat a_{+})/\sqrt{2}$. 
    Similarly $\hat b_{\pm}$ are mixed 
    to give outputs $\hat b_{\pm}^{'}¥$. The 
fields $a_{+}¥,b_{+}¥$ are coherent states of large amplitude.  
    The mixing is incorporated  
    so that both 
    fields, $\hat a_{\pm}^{'}¥$ say at $A$, incident on 
    the measuring apparatus are macroscopic. 
    The final measurements are made with the 
transformations (using polarisers or beam 
    splitters with variable transmission) 
    $c_{+}=\hat a_{+}^{'}\cos{\theta/2}+ \hat a_{-}^{'}\sin{\theta/2}$ 
and $c_{-}=\hat a_{+}^{'}\sin{\theta/2}- \hat a_{-}^{'}\cos{\theta/2}$, at $A$, 
and $d_{+}=\hat b_{+}^{'}\cos{\phi/2}+ \hat b_{-}^{'}\sin{\phi/2}$ 
and $d_{-}=\hat b_{+}^{'}\sin{\phi/2}- \hat b_{-}^{'}\cos{\phi/2}$, at 
$B$, followed by photodetection to give    
          $\hat S_{\theta}^A= (\hat c_+^\dagger \hat c_+-\hat c_-^\dagger \hat c_-)/2$ 
    and $\hat S_{\phi}^B= (\hat d_+^\dagger \hat d_+-\hat d_-^\dagger \hat 
    d_-)/2$.  The measurement is one of photon 
  number, and we define $\hat n_{\theta}^A = 2\hat S_{\theta}^A= 
  \hat c_+^\dagger \hat c_+-\hat c_-^\dagger \hat c_-$ 
    and $\hat n_{\phi}=2\hat S_{\phi}^B= 
    \hat d_+^\dagger \hat d_+-\hat d_-^\dagger \hat 
    d_-$.

\begin{figure}
\includegraphics[width=1.\textwidth]{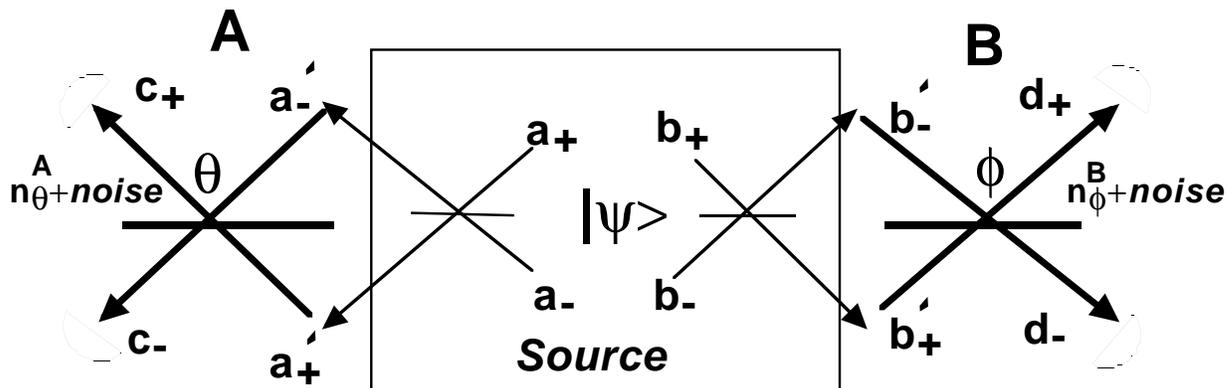}
\caption[diagrams]
{Schematic diagram of the 
 experimental arrangement used to give the two 
 different tests, described in sections 3 and 4, of macroscopic local 
 realism. The $a_+$ and $b_+$ represent strong coherent states. 
 For the macroscopic EPR experiment of section 3, the quantum state is chosen to ensure that the 
 output fields $a_{-}¥,b_{-}¥$ are EPR correlated with respect to 
 quadrature phase amplitudes.}%
\label{eps1}
\end{figure}

  In Figure 2 we demonstrate how the measurement $\hat S_{\theta}^A$ 
can be 
performed using an alternative arrangement, by introducing a relative phase shift 
$\theta$ and mixing with a $50/50$ beam splitter 
to produce $\hat c_{\pm}^{'}=\left( 
\hat a_{+} \pm \hat a_{-} \exp (-i\theta) \right)/\sqrt{2}$, followed by 
photodetection to give $\hat S_{\theta}^A= (\hat c_+^{'\dagger} 
\hat c_+^{'}¥-\hat c_-^{'\dagger} \hat c_-^{'}¥)/2$. 
It is to be clarified below that this measurement scheme 
corresponds to homodyne measurement~\cite{10} (used ~\cite{2,11} to detect
 subshot noise, or ``squeezed'', radiation) of the quadrature 
phase amplitudes of $a_{-}¥$ and $b_{-}¥$.

\begin{figure}
\includegraphics[width=1.\textwidth]{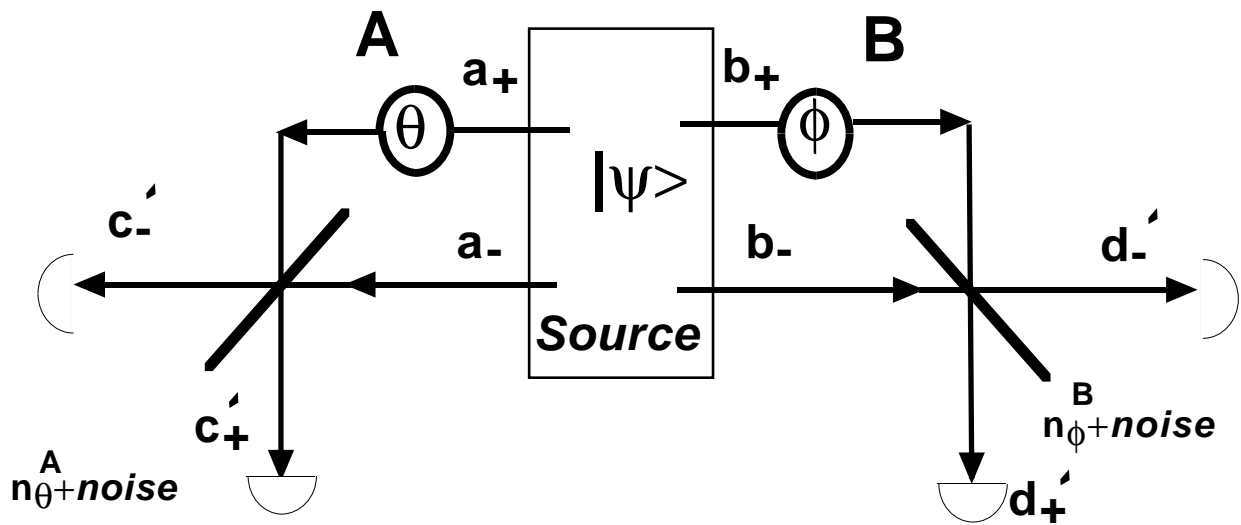}
\caption[diagrams]
{ An alternative arrangement used to measure $n_{\theta}^A$ and 
 $n_{\phi}^B$. 
 Fields $a_{+}¥,b_{+}¥$ are in coherent states of large amplitude 
  $\alpha=\beta=E$. This measurement scheme corresponds to 
  balanced homodyne 
 detection of the quadrature phase amplitudes $\hat X_{\theta}^A$ 
and $\hat X_{\phi}^B$ (defined in section 4), the $a_{+}¥,b_{+}¥$ being ``local 
oscillator'' fields.
 }%
\label{eps2}
\end{figure}

For certain systems the results for $\hat S_x^A$
and $\hat S_x^B$, and also $\hat S_y^A$
and $\hat S_y^B$,  are predicted by quantum mechanics
to be correlated, so that elements of reality 
may be deduced 
for the physical quantities $S_x^A$ and $S_y^A$ respectively. 
We 
consider the relevant uncertainty relation for $\hat S_x^A$ and $\hat S_y^A$~\cite{12}.  
\begin{equation}
\Delta \hat S_x^A \Delta \hat S_y^A\geq \left| \left\langle \hat 
S_z^A\right\rangle 
\right|/2= (\hat a_+^\dagger 
  \hat a_+-\hat a_-^\dagger \hat a_-)/4
\end{equation}
Fields $\hat a_+$ and $\hat b_+$ are 
coherent states $|\alpha>$ and  $|\beta>$ 
respectively with 
$\alpha=\beta=E$ real and large, so that 
$\hat a_+$ is replaced by c-number $E$
 giving 
\begin{eqnarray}
\hat S_x^A &=& E \left( \hat a_-+\hat a_-^{\dagger }\right)/2 =E \hat X_0^{A}¥/2 
\nonumber \\
\hat S_y^A &=& E \left( \hat a_--\hat a_-^{\dagger }\right) /2i=E 
\hat X_{\pi
/2}^{A}¥/2
\end{eqnarray}
where $\hat X_0^{A}¥=\hat a_-+\hat a_-^{\dagger }$ and $\hat 
X_{\pi /2}^{A}¥=\left( \hat a_--\hat a_-^{\dagger
}\right) /i$ are quadrature phase amplitudes. 
We denote
the indeterminacy in the ``element of reality'' $s_x^A$  by 
$\Delta _x$, and in the ``element of reality'' $s_y^A$
 by $\Delta _y$. In order to exclude
the possibility that the ``elements of reality'' cannot be described by a
quantum wavefunction, it is sufficient to establish (here $\left\langle \hat a_+^{\dagger
}\hat a_+\right\rangle >>\left\langle \hat a_-^{\dagger }\hat 
a_-\right\rangle$) 
\begin{equation}
\Delta _x\Delta _y<\left| \left\langle \hat S_z^A\right\rangle \right|/2 
=E^2/4
\end{equation}

The new feature of this EPR situation is the macroscopic
nature of the minimum uncertainty product: $2\left| \left\langle
\hat S_z^A\right\rangle \right| =E^2$ is a macroscopic (photon) number. 
The implication is that one need only use the premise of
``macroscopic local realism'', rather than local realism in its entirety,
 to arrive at the conclusion that quantum
mechanics is incomplete. With ``macroscopic local realism'',
one can only exclude the possibility of macroscopic
changes to the system at $A$, as a result of measurements made at $B$. 
One can predict with some error ($\Delta_1$ and $\Delta_2$ 
say for the $x$ and $y$ spin components respectively) the result of the
spin measurement at $A$. By macroscopic local realism, 
the spin is predetermined, but only to a
precision which excludes values macroscopically different to those in the
range predicted. The ``element of reality'' for the $x$ component of spin 
has a  range of
possible values, given by $\pm\Delta_x$ where $\Delta_x=\Delta_1 + \delta$ 
and $\delta $
is  microscopic or mesoscopic. The ``element of reality'' for 
the $y$ component has an 
indeterminacy $\pm \Delta_y$ where $\Delta_y= \Delta_2 + \delta$. 
This can still be sufficient to 
imply the EPR criterion (5) since 
the uncertainty limit $E^2$ is a macroscopic number. 
We only require for example 
 the differences 
$E/2-\Delta_1$ and $E/2-\Delta_2$ to be macroscopic numbers 
and we satisfy (5).

    To meet the situation of a macroscopic EPR experiment, where macroscopic 
local realism is used in the EPR conclusions, one needs 
 $\Delta _1\Delta _2<
\left| \left\langle \hat S_z^A\right\rangle \right|/2 
=E^2/4$
satisfied,  
but where $2\left| \left\langle \hat S_z^A\right\rangle \right| 
=E^2$, and $2(E/2-\Delta_1)$ and $2(E/2-\Delta_2)$ are macroscopic  
photon numbers. From the point 
of view of performing an experiment which is convincingly macroscopic, 
the preference would be to satisfy these criteria where   
measurement errors, and therefore also 
$\Delta _1$ and $\Delta _2$, are also large (photon) numbers.

So far we have assumed the existence of the required correlations for the 
spin operators. Such correlated
fields are predicted when $a_-,b_-$ are fields with
 ``EPR correlations'' for quadrature phase amplitudes, and 
$a_+,b_+$ are strong coherent
fields. EPR correlations for quadrature phase amplitudes were  
defined in~\cite{12}, and have received much interest as  
fields enabling quantum teleportation for continuous 
variables~\cite{13}. 
EPR fields may be generated by the parametric interaction $H=i\hbar 
\kappa\left( \hat a_-^{\dagger }\hat b_-^{\dagger }-\hat a_-\hat b_-\right)$, 
where here $\kappa$ represents the strength of nonlinear interaction. 
 This two-mode squeezed light represents the logical choice for the quantum state 
$|\phi>$ in figure 1.
(It is possible 
to generate the required EPR correlations from a single-mode squeezed 
state passed using a $50/50$ beam splitter~\cite{12,13}.) 

Using the parametric example, we consider the quadrature phase  
amplitudes $\hat X_0^{A}¥,\hat X_{\pi /2}^{A}¥$, and $\hat 
X_0^{B}¥$, $\hat X_{\pi /2}^{B}¥$ where 
$\hat X_0^{B}¥=b_-+b_-^{\dagger }$ and $\hat X_{\pi /2}^{B}¥=\left( 
\hat b_--\hat b_-^{\dagger
}\right) /i$.  
With vacuum inputs, the output solutions after a  
time $t$, for  
$\kappa t \rightarrow \infty$, satisfy  
$\hat X_0^{A}¥\left( t\right)=\hat X_0^{B}¥\left(
t\right) $ and $\hat X_{\pi /2}^{A}¥\left( t\right)=-\hat 
X_{\pi /2}^{B}¥\left( t\right) $,
indicating a maximum correlation$^{\left[ 13\right] }$ between the 
results of measurements $
\hat X_0^{A}¥\left( t\right) $ and $\hat X_0^{B}¥\left( t\right)$, and also 
$\hat X_{\pi /2}^{A}¥\left(
t\right) $ and $\hat X_{\pi /2}^{B}¥\left( t\right).$ The 
$\hat a_+,\hat b_+$ fields are coherent states $\left| \alpha
\right\rangle $ and $\left| \beta\right\rangle $ of large intensity
 so that $\hat S_x^A=\alpha \hat 
X_0^{A}¥(t)/2$, $\hat S
_x^B=\beta\hat X_0^{B}¥(t)/2$, $\hat S_y^A=\alpha \hat 
X_{\pi /2}^{A}¥(t)/2$ and 
$\hat S_y^B=\beta\hat X_{\pi /2}^{B}¥(t)/2$, and we have the required 
correlations. The reader is referred to articles~\cite{2,12,9} 
for information on the evaluation of $\Delta
_1$ and $\Delta _2$.

The macroscopic EPR experiment performed with results in accordance with quantum 
mechanics would logically lead to the conclusion that: either macroscopic local 
realism is invalid; or that quantum mechanics is incomplete. EPR 
experiments have perhaps not yet been widely considered important in their own 
right, since previously they have been based on  
local realism, a 
premise dismissed by Bell inequality experiments. 
(Actually this is not quite correct, since detector inefficiencies 
have prevented a true violation of a Bell inequality. Proposed EPR 
quadrature experiments do not suffer this problem).  In our 
macroscopic example this is not correct. The validity of macroscopic local 
realism has not been tested. 

The macroscopic EPR experiment has been performed in a partly 
satisfactory way by Ou et al~\cite{2}, with the arrangement of Figure 2. 
For a conclusive result the arrangement of Figure 1 is 
preferred. The experimental scheme of Ou 
et al  
suffers the disadvantage 
that fields $a_{-}¥,b_{-}¥$ are microscopic. 
This is irrelevant in that the quantity 
measured, and to 
which the elements of reality relate, is not $\hat X_{\theta}^{A}$  
but $\hat S_{\theta}^{A}$. 
Nevertheless, the microscopic nature of $\hat a_{-}¥,\hat b_{-}¥$ incident on 
the  
measurement apparatus gives the impression of a microscopic experiment. 
The scheme of Figure 1 where both fields incident on the measurement 
apparatus are  
macroscopic is more transparent, making it clear that the ``local 
oscillator'' fields $a_{+},b_{+}$ form part of the system.   
 It  
is also essential to ensure measurement events (the 
selection of $\theta$ or $\phi$) at $A$ and $B$ causally separated, 
as in delayed-choice Bell inequality experiments~\cite{4}.  Since a wide variety of 
squeezing experiments have been performed, including squeezing in 
pulsed fields, an experimental realisation of the 
macroscopic EPR experiment would seem very feasible.

   \section{A Direct Test of Macroscopic Local Realism: Bell Inequalities based on Macroscopic Local Realism}

  We prove that for coarse measurements with macroscopic 
   uncertainties, Bell inequalities can be derived 
   using only the premise of macroscopic local realism~\cite{14}.
    Our proposed experiment 
    is again depicted in Figure 1,, except 
that the quantum state $|\phi>$ will be chosen differently. The 
parametric interaction used as a source for EPR correlations 
in the Ou et al experiment is not directly siutable for this 
experiment, as in this case a positive Wigner function exists.  
This positive Wigner function can act as a local hidden 
variable theory to describe the quantum predictions [3], and thus 
prevent a violation of a Bell inequality as we derive here. With an 
appropriate choice of quantum state then, we measure simultaneously.
  
    We measure simultaneously at 
$A$ and $B$ the Schwinger  
     operators 
          $\hat S_{\theta}^A$ and $\hat S_{\phi}^B$. 
    The result for the photon number differences $
    \hat n_{\theta}^{A}$ and  
    $\hat n_{\phi}^{B}$ is of the form $n+noise$, where $n$ is the result 
    in the absence of the noise. 
    We introduce a random noise function (gaussian distribution of standard 
 deviation $\sigma$) at 
  each of $A$ and $B$, and define probabilities such as  
  $P^A(noise\geq x)$, that the $noise$ at $A$ 
   is greater than or equal to the 
  value $x$. 
 
    The results of measurements are classified as $+1$ if the photon number 
    difference   
  is positive or zero, and $-1$ otherwise. 
  We determine the 
following probability distributions: $P_{+}^{A}(\theta)$ 
for obtaining $+$ at $A$; $P_{+}^{B}(\phi)$ for obtaining $+$ at $B$; and 
$P_{++}^{AB}(\theta,\phi)$ the joint probability of $+$ at 
both $A$ and $B$.

 We define the  
 probability $P_{ij}^{0,AB}(\theta,\phi)$ for 
 obtaining results $i$ and 
 $j$ respectively upon joint measurement of 
 $\hat n_{\theta}^A$ at $A$, and $\hat n_{\phi}^B$ at $B$, 
  in the absence of the applied noise $\sigma$. With noise present, measured 
 probabilities become  
   \begin{equation}
   P_{++}^{AB}(\theta,\phi) = \sum_{i,j=-\infty}^{\infty}
   P_{ij}^{0,AB}(\theta,\phi) P^A(noise\geq -i) P^B(noise\geq -j)
\end{equation}

   Local realism as defined by 
  Einstein-Podolsky-Rosen, Bell and 
  Clauser-Horne~\cite{1} implies the well known expression.  
    \begin{equation}
   P_{ij}^{0,AB}(\theta,\phi) = \int \rho(\lambda) \quad p_{i}^A(\theta, \lambda ) 
   p_{j}^B(\phi, \lambda )\quad d\lambda  
	\label{8}
\end{equation}
Local realism implies a set of elements of reality, or 
hidden variables $\lambda$ (with probability distribution $\rho(\lambda)$),  
not specified by quantum theory. For our experiment, a precise prediction of $\hat n_{\theta}^{A}$ 
is not possible given a measurement at $B$, for any choice $\phi$ at 
$B$. The elements of reality then do not take on definite values and 
local realism is only sufficient to imply 
a probability $p_{i}^A(\theta, \lambda )$ for 
 the result $i$ of the measurement $\hat n_{\theta}^{A}$, for a given $\lambda$. 
  The independence of  $p_i^A (\theta, \lambda)$ 
  on $\phi$ is based on the locality assumption. 
   
  With macroscopic local realism the locality condition is relaxed, but only up to the level of 
  $M$ photons, where $M$ is not macroscopic, by maintaining 
  that the measurement at $B$ cannot instantaneously 
  change the result at $A$ by an amount exceeding $M$ photons.
 Where our predicted result 
  at $A$ is 
  $i^{'}¥$ using local realism, macroscopic local realism 
  allows the result to be $i=i^{'}¥+m_{A}¥$ where $m_{A}¥$ can be any 
  number not macroscopic. Importantly, while $i^{'}¥$ is not dependent on 
  the choice $\phi$ at $B$, the value $m_{A}¥$ which is 
  not macroscopic can be.
   The macroscopic local realism assumption is that 
   the conditional probability  
   $p_{i}^A(\theta, \lambda )$ in equation (7) is expressible as the 
   convolution: 
       \begin{equation}
     p_{i}^A({\theta,\phi}, \lambda )= \sum_{m_A=-M}^{+M}
      p_{m_A}^{A} (i^{'},{\theta,\phi},\lambda) 
      p_{i'=i-m_A}^{A} (\theta, \lambda ). 
	\label{9}
\end{equation}
The original local probability $p_{i'}^{A} (\theta, \lambda )$ 
can be convolved with a microscopic 
nonlocal probability function $p_{m_A}^{A} (i^{'},{\theta,\phi},\lambda)$, 
the only restriction being that the nonlocal distribution does not 
provide macroscopic perturbations, so that the 
	probability of getting a nonlocal change outside the range $m_A=-M,...,
	+M$ is zero. 
	Equivalently we must have 
\begin{equation}	
	 \sum_{m_A=-M}^{M}
       p_{m_A}^{A} (i^{'},{\theta,\phi},\lambda)  = 1.   
  \end{equation}   
We substitute the macroscopic locality 
assumption (8) into (7) to obtain the 
prediction for the measured probabilities (6).
Recalling $i=i'+m_A$, $j=j'+m_B$ we change the $i$, $j$ summation to 
one over $i'$, $j'$ to get
\begin{eqnarray}
P_{++}^{AB}(\theta,\phi) &=& 
   \sum_{i',j'=-\infty}^{\infty}\int \rho(\lambda)p_{i'}^{A} 
   (\theta,\lambda) \nonumber \\ 
    &\times& \biggl[\sum_{m_A=-M}^{M}
        p_{m_A}^{A} (i^{'},{\theta,\phi},\lambda)  p_{j'}^{B} (\phi, \lambda)
       P^A(noise\geq -(i'+m_A)) \biggr]  \biggr. \nonumber \\
 &\times&  \biggl[\sum_{m_B=-M}^{M}
\biggl.     p_{m_B}^{B} (j^{'},{\phi,\theta},\lambda) 
     P^B(noise\geq -(j'+m_B)) \biggr] d\lambda  
 	\label{12}
\end{eqnarray}
We assume that 
 the noise function $noise$ is slowly varying over the 
microscopic (or mesoscopic) range $-m_{A},..+m_{A}$ for 
which nonlocal perturbations are 
possible according to macroscopic local realism:   
 \begin{eqnarray}
\sum_{m_A=-M}^{M}
        p_{m_A}^{A} (i^{'},{\theta,\phi},\lambda) 
       P^A(noise\geq -(i'+m_A)) \nonumber \\
       \approx P^A(noise\geq -i') \sum_{m_A=-M}^{M}
        p_{m_A}^{A} (i^{'},{\theta,\phi},\lambda).
   \end{eqnarray}      
This is only valid 
if $\sigma$ is macroscopic.
Using (11), one simplifies to get the final form $
   P_{++}^{AB}(\theta,\phi)=\sum_{i',j'}\ \int \rho(\lambda)
     p_{i'}^{A} (\theta, \lambda )  p_{j'}^{B} (\phi, \lambda ) d\lambda 
    \times P^A(noise\geq -i') P^B(noise\geq -j')$. 
  This prediction of the hidden variable 
	theory is now given in a (local) form like that of (7), from which Bell-
	Clauser-Horne  
	inequalities~\cite{3} follow, for example:
	  \begin{equation}
 S={{P_{++}^{AB}(\theta,\phi)-P_{++}^{AB}(\theta,\phi')+P_{++}^{AB}(\theta',\phi)
 +P_{++}^{AB}(\theta',\phi')}\over{P_{+}^{A}(\theta')+P_{+}^{B}(\phi)}} \leq 1.
	\label{eqnbell}
\end{equation}

 Violation of Bell inequalities (12) with macroscopic noise terms 
	($\sigma$ macroscopic) 
	 is evidence of a failure of macroscopic local 
	realism.
	We propose a quantum state with this property ($I_{0}$ is a modified Bessel 
    function, $r_{0}=1.1$).  
  \begin{equation}
	|\psi\rangle = [I_{0}(2r_{0}^{2})]^{-1/2}	|\alpha>_{a_{+}} |\beta>_{b_{+}}
	\left(\sum_{n=0}^{\infty}
	\frac{(r_{0}^{2})^{n}}{n!}|{n}>_{a_{-}} |{n}>_{b_{-}}\right). 
\end{equation}  
$|\alpha>_{a_{+}¥}$ and  $|\beta>_{b_{+}¥}$ are coherent states with 
$\alpha$, $\beta$ real and large. $|{n}>_{k}$ is a Fock 
state for field $k$. The fields $\hat 
a_-$ 
and $\hat b_-$ are microscopic and are generated 
in a pair-coherent state~\cite{15}. 
 The quantum prediction for (13) is shown in Figure 3.
Violations of the Bell inequality (12) in the absence 
 of $noise$ are shown in curve (a).
Violations are still  
 possible (curve (b)) in the presence of increasingly 
 larger absolute noise $\sigma$, simply by 
 increasing $\alpha=\beta=E$.
This violation of the Bell inequality (12) with macroscopic noise 
$\sigma$ implies the failure of macroscopic local 
	realism.
\begin{figure}
 
\includegraphics[scale=.8]{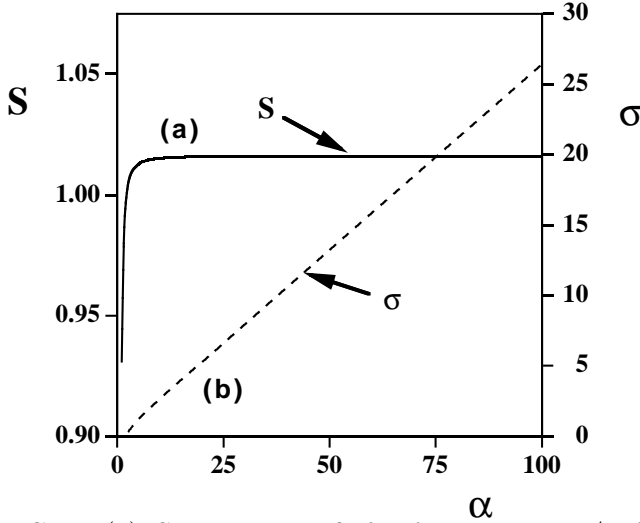}
\caption[Violations of the Bell's inequality found 
for the quantum state (14).]
{(a) $S$ versus $\alpha=\beta$, for 
$\theta=0,\phi=-\pi/4,\theta'=\pi/2,\phi'=-3\pi/4$ for the 
quantum state (13) with no noise present.
(b) Maximum noise $\sigma$ still giving a violation of the Bell 
inequality (12), versus $\alpha$.
}% 
\label{eps3}
\end{figure}

The large $\alpha$,$\beta$ limit is 
crucial in determining whether the violation of macroscopic local 
realism will occur. We see from (1) and (2) that (letting 
$\alpha=\beta=E$) 
$\hat S_{\theta}^A=E \hat X_{\theta}^A/2$ and  
$\hat S_{\phi}^B=E \hat X_{\phi}^{B}/2$, 
where  $\hat X_{\theta}^A=\hat a_-exp(-i\theta)+\hat a_-^\dagger exp(i\theta)$ 
and $\hat X_{\phi}^B=\hat b_-exp(-i\phi)+\hat b_-^\dagger exp(i\phi)$ are 
the quadrature phase amplitudes of fields $\hat a_{-}¥$ and $\hat b_{-}¥$. 
Violations of Bell inequalities for    
measurements $\hat X_{\theta}^A$, $\hat X_{\phi}^B$ on state (13) 
have recently been predicted~\cite{16}, confirming Figure 3(a). 
 It is always the case that such violations of a Bell inequality will 
vanish  
 when 
gaussian noise of sufficiently large standard deviation $\sigma_{0}$ is added to 
the measurements $\hat X_{\theta}^A, \hat X_{\phi}^B$.  
With $\alpha=E$ sufficiently large, this corresponds to a macroscopic noise value of $E \sigma_{0}¥$ in the photon 
number measurement $\hat n_{\theta}^A$. 
Therefore any state $|\psi\rangle$ which shows a failure of local realism for 
measurements  $\hat X_{\theta}^A$ 
and $\hat X_{\phi}^B$ will also indicate a 
violation of macroscopic local realism, provided $\alpha$, $\beta$ are 
large. This is relevant since other such states have been recently 
predicted~\cite{17}, such as an odd or even coherent state passed through 
a beam splitter and parametric interaction~\cite{16}.

This Bell inequality test is 
logically more straightforward than the EPR test, and is 
stronger, potentially leading to the rejection of macroscopic local realism 
outright. Appropriate states however are likely to be  
difficult to prepare.  Unlike the EPR 
test it becomes strictly necessary to ensure the measurement 
uncertainty in photon number is macroscopic in absolute terms, 
because of assumption (11).


\begin{references}
\small
%
\bibitem{1}	A. Einstein, B. Podolsky and N. Rosen, Phys. Rev. 
\textbf{47}, 777 (1935).
%
\bibitem{2} Z. Y. Ou, S. F. Pereira, H. J. Kimble and K. C. Peng, 
Phys. Rev. Lett. \textbf{68}, 3663 (1992). 
 See also recent experiments of 
Yun Zhang, hai Wang, Xiaoying Li,Jietai Jing, Changde Xie and Kunchi 
Peng, Phys. Rev. A{\bf 62},023813(2000); Ch. Silberhorn, P. K. Lam, G. 
Wasik, N. Korolkova and G. Leuchs, presented at Europe IQEC (2000).
%
\bibitem{3} J. S. Bell, Physics, \textbf{1}, 195 (1965). 
J. F. Clauser and A. Shimony, 
Rep. Prog. Phys. \textbf{41}, 1881 (1978). D. M. Greenberger, 
M. Horne and A. Zeilinger. In: {\itshape Bell's 
Theorem, Quantum Theory and Conceptions of the Universe}, ed. by 
M.Kafatos  
(Kluwer, Dordrecht, The Netherlands 1989), p. 69.
 %
\bibitem{4}	A. Aspect,  P. Grangier and G. Roger,  Phys. Rev. Lett. 
\textbf{49},  91  (1982). 
A. Aspect,  J. 
Dalibard and G. Roger,  ibid. \textbf{49},  1804  (1982). W. Gregor, T. 
Jennewein and A. Zeilinger, Phys. Rev. Lett. \textbf{81}, 5039 (1998).
%
\bibitem{5} 	N. D. Mermin, Phys. Rev. D \textbf{22},  356 (1980).
 P. D. Drummond, Phys. Rev. Lett. \textbf{50}, 1407 (1983). 
 A. Garg and 
N. D. Mermin, Phys. Rev. Lett. \textbf{49}, 901 (1982).
 S. M. Roy and V. Singh, Phys. 
 Rev. Lett. \textbf{67}, 2761 (1991). A. Peres, Phys. Rev. A \textbf{46}, 4413 
 (1992).
 M. D. Reid and W. J. 
 Munro, Phys. Rev. Lett. \textbf{69}, 997 (1992). 
 G. S. Agarwal, Phys. Rev. A \textbf{47}, 4608 (1993). 
  D. Home and A. S. 
 Majumdar, Phys. Rev. A \textbf{52}, 4959 (1995). W. J. Munro 
 and M. D. Reid, Phys. Rev. A \textbf{47}, 4412 (1993). C. Gerry, Phys. Rev. A 
 \textbf{54}, 2529, (1996). 
N. D. Mermin, Phys. Rev. Lett. \textbf{65}, 1838 (1990). B. J. Sanders, Phys. Rev. A\textbf{45}, 6811, 
(1992).
%
\bibitem{6} E. Schr\"odinger,  Naturwissenschaften \textbf{23}, 812 (1935). 
%
\bibitem{7}  C. Monroe, D. M. Meekhof, B. E. King and D. J. Wineland, 
Science, \textbf{272},1131 (1996).  
M. Brune, E. Hagley, J. Dreyer, X. Maitre, A. Maali, C. 
Wunderlich, J. M. Raimond and S. Haroche, Phys. Rev. Lett.\textbf{77}, 4887
 (1996).
M. W. Noel and C. R. Stroud, Phys. Rev. Lett.\textbf{77},1913 (1996).
J. R. Friedman, V. Patel, W. Chen, S. K. Tolpygo and J. E. 
Lukens, Nature{\bf 406}, 43-45 (2000). 
%
\bibitem{8}  A. J. Leggett and A. Garg,  Phys. Rev. Lett. \textbf{54}, 857 (1985). 
%
\bibitem{9} M. D. Reid, Europhys. Lett. \textbf{36}, 1 (1996). M. D. 
Reid,  
Quantum Semiclass. Opt.\textbf{9},489 (1997). M. D. Reid and P. Deuar, 
Ann. Phys. \textbf{265}, 52 (1998). 
%
\bibitem{10}  H. P. Yuen and V. W. S. Chan, Opt. Lett. \textbf{8}, 177 (1983).
%
\bibitem{11} M. D. Levenson, R. M. Shelby, M. D. Reid and D. F. Walls, 
Phys. Rev. Lett. \textbf{57}, 2473 (1986).
 %
\bibitem{12} M. D. Reid, Phys. Rev. A \textbf{40}, 913 (1989).
%
\bibitem{13} A. Furasawa, J. Sorensen, S. Braunstein, C. Fuchs, H. 
Kimble and E. Polzik, Science {\bf 282}, 706 (1998). 
 L. Vaidman, Phys. Rev. A{\bf 49},1473 (1994). 
S. Braunstein and H. J. Kimble, Phys. Rev. Lett. {\bf 
80}, 869 (1998).
%
\bibitem{14} M. D. Reid, Phys. Rev. Lett.{\bf 84}, 2765 (2000);
Phys. Rev. A. {\bf 62} 022110 (2000). 
%
\bibitem{15} G. S. Agarwal, Phys. Rev. Lett.\textbf{57},827, (1986). 
M. D. Reid and L. Krippner, Phys. Rev. A\textbf{47}, 552 (1993).         
 %
\bibitem{16} A. Gilchrist, P. Deuar and M. D. Reid, Phys. Rev. 
 Lett. \textbf{80}, 3169 (1998); Phys. Rev. A\textbf{60},4259 (1999). 
 %
\bibitem{17}   
B. Yurke, M. Hillery and D. Stoler, Phys. Rev. A {\bf 60}, 3444 
 (1999). 
 W. J. Munro and G. J. Milburn, Phys. Rev. Lett. \textbf{
 81}, 4285 (1998). W. J. Munro, Phys. Rev. A\textbf{59}, 4197 (1999).
%
\bibitem{18} A. Peres, Found. Phys.\bf{22}, 819 (1992).
%
\end{references}
\end{document}